\documentclass[letterpaper,aps,prl,onecolumn,notitlepage,nofootinbib,10pt,superscriptaddress,longbibliography]{revtex4-1}
\usepackage[ruled,vlined]{algorithm2e}
\usepackage{amsmath}
\usepackage{amssymb}
\usepackage{amsfonts}
\usepackage{amsthm}
\usepackage{youngtab}

\usepackage{graphicx}

\usepackage{color}
\definecolor{darkblue}{rgb}{0.,0.,0.4}
\definecolor{darkred}{rgb}{0.5,0.,0.}
\usepackage[pdftex,colorlinks=true,linkcolor=darkblue,citecolor=darkred,urlcolor=darkblue]{hyperref}

\newtheorem{thm}{Theorem}
\newtheorem*{thm*}{Theorem}

\theoremstyle{definition}

\newtheorem{observation}{Observation}

\theoremstyle{plain}
\newtheorem{prob}{Problem}
\makeatletter
\newtheorem*{rep@theorem}{\rep@title}
\newcommand{\newreptheorem}[2]{%
\newenvironment{rep#1}[1]{%
 \def\rep@title{#2 \ref{##1} (restatement)}%
 \begin{rep@theorem}}%
 {\end{rep@theorem}}}
 \SetKwInput{Input}{Input}
\SetKwInput{Output}{Output}
\usepackage{lipsum}
\def\C{\mathbb{C}}
\makeatother

\newreptheorem{thm}{Theorem}

\newcommand{\op}[2]{|#1\rangle \langle #2|}
\newcommand{\nc}{\newcommand}
\nc{\rnc}{\renewcommand}

\nc\eps{\epsilon}

\nc\bbC{\mathbb{C}}

\nc\bbF{\mathbb{F}}
\nc\bbM{\mathbb{M}}
\nc\bbN{\mathbb{N}}
\nc\bbR{\mathbb{R}}
\nc\bbS{\mathbb{S}}
\nc\bbZ{\mathbb{Z}}

\nc\bp{\mathbf{p}}
\nc\bq{\mathbf{q}}

\nc\benum{\begin{enumerate}}
\nc\eenum{\end{enumerate}}
\nc\bit{\begin{itemize}}
\nc\eit{\end{itemize}}

\nc{\todo}[1]{\textcolor{red}{todo: #1}}

\nc{\Anote}[1]{\textcolor{red}{Aram note: #1}}
\nc{\Ynote}[1]{\textcolor{red}{Nengkun note: #1}}

\nc\cA{\mathcal{A}}
\nc\cB{\mathcal{B}}
\nc\cC{\mathcal{C}}
\nc\cD{\mathcal{D}}
\nc\cE{\mathcal{E}}
\nc\cF{\mathcal{F}}
\nc\cG{\mathcal{G}}
\nc\cH{\mathcal{H}}
\nc\cI{\mathcal{I}}
\nc\cJ{\mathcal{J}}
\nc\cK{\mathcal{K}}
\nc\cL{\mathcal{L}}
\nc\cM{\mathcal{M}}
\nc\cN{\mathcal{N}}
\nc\cO{\mathcal{O}}
\nc\cP{\mathcal{P}}
\nc\cQ{\mathcal{Q}}
\nc\cR{\mathcal{R}}
\nc\cS{\mathcal{S}}
\nc\cT{\mathcal{T}}
\nc\cU{\mathcal{U}}
\nc\cV{\mathcal{V}}
\nc\cW{\mathcal{W}}
\nc\cX{\mathcal{X}}
\nc\cY{\mathcal{Y}}
\nc\cZ{\mathcal{Z}}

\def\be#1\ee{\begin{equation}#1\end{equation}}
\def\bea#1\eea{\begin{eqnarray}#1\end{eqnarray}}
\def\beas#1\eeas{\begin{eqnarray*}#1\end{eqnarray*}}
\def\ba#1\ea{\begin{align}#1\end{align}}
\def\bas#1\eas{\begin{align*}#1\end{align*}}
\def\bpm#1\epm{\begin{pmatrix}#1\end{pmatrix}}

\rnc\L{\left}
\nc\R{\right}
\nc\ra{\rightarrow}
\nc\ot{\otimes}

\begin{document}

\title{Sample ``optimal” Quantum identity testing via Pauli Measurements}

\author{Nengkun Yu}

\affiliation{Centre for Quantum Software and Information, Faculty of Engineering and Information Technology, University of Technology, Sydney, NSW 2007, Australia}

\date{\today}


\begin{abstract}
In this paper, we show that $\Theta(\mathrm{poly}(n)\cdot\frac{4^n}{\epsilon^2})$ is the sample complexity of testing whether two $n$-qubit quantum states $\rho$ and $\sigma$ are identical or $\epsilon$-far in trace distance using two-outcome Pauli measurements.
\end{abstract}


\maketitle
\section{Introduction}
At this stage of development of quantum computation, testing the properties of new devices is of fundamental importance \cite{MdW13}. Quantum state tomography is to decide how many copies of an unknown mixed quantum state $\rho\in\mathcal{D}(\C^d)$ is necessary and sufficient to output a good approximation of ${\rho}$ in trace distance, with high probability. A sequence of work \cite{compressed,FlammiaGrossLiuEtAl2012,Vlad13,KRT14,HHJ+16} is showed that $\Theta(\frac{d^3}{\epsilon^2})$ copies are sufficient and sufficient for quantum state tomography in a trace distance of no more than $\epsilon$, by performing measurements on each copy of $\rho$ independently. If one is allowed to measure many copies of $\rho$ simultaneously, the sample complexity of state tomography is $\Theta(\frac{d^2}{\epsilon^2})$ in \cite{HHJ+16,OW16,OW17}.

As one of the most fundamental problem in the general framework of quantum state property testing, the problem ``quantum identity testing'' has received much attention: Given many copies of unknown quantum states $\rho$ and $\sigma$, and the goal is to distinguish the case that they are identical and that they are $\epsilon$-far in trace distance. Here the goal is to derive a measurement scheme to learn some non-trivial property, using as few as possible copies. 

This problem is a quantum analog of the identity testing of probabilistic distributions, a central problem of distribution testing. The idea of identity testing has been extensively explored in studying other property testing problems \cite{BFR+00,Batu:2001:TRV:874063.875541,GR11,Pan08,VV11,Valiant:2014:AIP:2706700.2707449,Chan:2014:OAT:2634074.2634162,DK16}. 

For practical purposes, the results from cases where $\sigma$ is a known pure state have been extensively studied in the independent measurement setting \cite{PhysRevLett.106.230501,PhysRevLett.107.210404,AGKE15}. 
For the general mixed states case, \cite{OW15} solved the problem, in the joint measurement setting, where $\sigma$ is a maximally mixed state case by showing that $\Theta(\frac{d}{\epsilon^2})$ copies are necessary and sufficient. Importantly, the sample complexity of the general problem was proven to be the same in \cite{BOW17}. In \cite{yu2019quantum}, we provide an independent measurement scheme using $\Theta(\frac{d^2}{\epsilon^2})$ copies for this problem. Interestingly, \cite{bubeck2020entanglement} showed that the sample complexity of this problem in the independent measurement setting, where $\sigma$ is a maximally mixed state case, is $\Theta(\frac{d^{3/2}}{\epsilon^2})$ with nonadaptive measurements, and $\Omega(\frac{d^{4/3}}{\epsilon^2})$ with adaptive measurements.

To achieve the optimal complexity in the general mixed state scenario, \cite{HHJ+16,OW16,OW17,OW15,BOW17} empolys highly entangled measurements. The other side of these entangled measurements is the difficulty in the implementation: all $\Theta(\frac{d^{1,2}}{\epsilon^2})$ copies must be stored in a noiseless environment; for $n$-qubit system, the needed measurement is within dimension $d^{\frac{d^{1,2}}{\epsilon^2}}=2^{\Omega(n\cdot 2^{n,2n}})$, a double exponential function of the number of qubits. The complexity $\Theta(\frac{d^{3/2}}{\epsilon^2})$ for independent measurement of \cite{bubeck2020entanglement} requires $n$-qubit random unitary, which is highly entangled in $n$-qubit system.

This observation leads to the question: What if entangled measurements are not allowed? As the most important class of quantum measurements, Pauli measurements are of great interest.

\subsection{Our results}
The following \textit{quantum identity testing} problem was extensively studied in \cite{OW15,BOW17, bubeck2020entanglement}
\begin{prob} \label{identity}
Given two unknown $n$-qubit quantum mixed states $\rho$ and $\sigma$, and $\epsilon>0$,
the goal is to distinguish the two cases
\begin{align}
||\rho-\sigma||_1>\epsilon
\end{align}
and 
\begin{align}
\rho=\sigma
\end{align}
How many copies of $\rho$ and $\sigma$ are needed to achieve this goal, with high probability? 
\end{prob}
In this paper, we focus on the sample complexity of this problem using Pauli measurements. In particular, we show that
\begin{thm}
 The sample complexity of the \textit{quantum identity testing} problem is ${\Theta}(\frac{\mathrm{poly}(n)\cdot 4^n}{\epsilon^2})$ using two-outcome Pauli measurements.
 \end{thm}

\section{Preliminaries}

A positive-operator valued measure (POVM) on a finite dimensional Hilbert space $\mathcal{H}$ is a set of positive semi-definite matrices $\mathcal{M}=\{M_i\}$ such that
\begin{align*}
\sum M_i=I_{\mathcal{H}}.
\end{align*}

We use $\sigma_I$, $\sigma_X,\sigma_Y$ and $\sigma_Z$ to denote Pauli matrices,
\begin{align*}
\sigma_I=\begin{bmatrix}1 &0\\0&1\end{bmatrix}, \sigma_X=\begin{bmatrix}0 &1\\1&0\end{bmatrix},  \sigma_Z=\begin{bmatrix}1 &0\\0&-1\end{bmatrix}, \sigma_Y=\begin{bmatrix}0 &i\\-i&0\end{bmatrix}.
\end{align*}

For $U\in\{\sigma_X,\sigma_Y,\sigma_Z\}$ with $U=\op{\psi_0}{\psi_0}-\op{\psi_1}{\psi_1}$, we use Pauli measurement to denote the following measurement,
\begin{align*}
M_0=\op{\psi_0}{\psi_0}, M_1=\op{\psi_1}{\psi_1}.
\end{align*}

The following is widely known,
\begin{observation}
For two unknown binary distributions $p$ and $q$, the sample complexity of distinguishing the two cases $p=q$ and $||p-q||_2>\epsilon$ with probability at least $1-\delta$ is $\Theta(\frac{\log(1-\delta)}{\epsilon^2})$.
\end{observation}
There is no difference for $2$-norm (the Euclidean norm) and $1$-norm (total variation) in binary distribution setting.

\section{Quantum Identity Testing}

In this section, we provide a measurement scheme which uses $\mathcal{O}(\frac{n^4\cdot 4^n}{\epsilon^2})$ copies to solve Problem \ref{identity}. The main tool is an algorithm for testing the property of probability collections.

\subsection{Upper bound}
The property testing of collections of discrete distributions was studied in \cite{LRR13,DK16}. In this section, we study the identity testing of discrete distributions in the query model. Suppose we are given two collection of binary distributions in a query model, with $m$ binary distributions $p_1$, . . . , $p_m$ for the first collection and $q_1$, . . . , $q_m$ for the second collection such that for any given index $i$ that we can choose to access $p_i$ and $q_i$ and obtain samples. The goal is to distinguish the case of $p_i=q_i$ for all $i$ from the case
$$
\frac{1}{m}\sum_{i=1}^m ||p_i-q_i||_2^2>\epsilon^2
$$
\begin{algorithm}[H]
	\Input{Access to binary distributions $p_1$, . . . , $p_m$ and $q_1$, . . . , $q_m$ with $ \epsilon> 0$.}	
    \Output{"Yes" with a probability of at least $\frac{2}{3}$ if $p_i=q_i$ for all $1\leq i\leq m$; and "No" with a probability of at least $\frac{2}{3}$ if $
\frac{1}{m}\sum_{i=1}^m ||p_i-q_i||_2^2>\epsilon^2
$.}
    Let $L$ be a sufficiently large constant\;
    \For{$k^2\gets 0$ \KwTo $\lceil\log_2 m\rceil$}{
     \ \ Select $2^{k}\cdot ({k^2+1})\cdot L$ uniformly random elements $1\leq i\leq m$\;
    \ \ For each selected $i$, distinguish between $p_i=q_i$ and $||p_i-q_i||_2^2>2^{k-1}\epsilon^2$ with a failure probability of at most $L^{-2}6^{-k}$\;
   \ \  If any of these testers returned "No", return "No"\;
}
Return "Yes"\;
	\caption{\textsf{An Identity Test for Collections with a Query Model}}
	\label{algo:Identity-Test-Collection-Query-Model}
\end{algorithm}

If $p_i=q_i$ for all $1\leq i\leq m$, then the probability of success is at least 
\begin{align*}
1-\sum_{k=0}^{\lceil\log_2 m\rceil} (1-p_k),
\end{align*}
where
\begin{align*}
p_k\geq  (1-2^{k} \cdot (k^2+1)\cdot L\cdot L^{-2}6^{-k})=1-\frac{k^2+1}{3^kL}.
\end{align*}
Then
\begin{align*}
1-\sum_{k=0}^{\lceil\log_2 m\rceil} (1-p_k)\geq 1-\sum_{k=0}^{\lceil\log_2 m\rceil}\frac{k^2+1}{3^kL}\geq 1-\mathcal{O}(\frac{1}{L}).
\end{align*}
On the other hand, if $\frac{1}{m}\sum_{i=1}^m ||p_i-q_i||_2^2>\epsilon^2$, we observe the following
\begin{align*}
\frac{1}{m}\sum_{i=1}^m ||p_i-q_i||_2^2<& \frac{|\{i: ||p_i-q_i||_2^2<\frac{\epsilon^2}{2}\}|\frac{\epsilon^2}{2}}{m}+\sum_{k=0} \frac{|\{i:2^{k-1}\epsilon^2\leq ||p_i-q_i||_2^2< 2^k\epsilon^2\}|{2^k\epsilon^2}}{m}\\
<&\frac{\epsilon^2}{2}+\sum_{k=0} \frac{|\{i:2^{k-1}\epsilon^2\leq ||p_i-q_i||_2^2\}|{2^k\epsilon^2}}{m}
\end{align*}
Thus,
\begin{align*}
\frac{\epsilon^2}{2}+\sum_{k=0} \frac{|\{i:2^{k-1}\epsilon^2\leq ||p_i-q_i||_2^2\}|{2^k\epsilon^2}}{m}
>\epsilon^2>\frac{\epsilon^2}{2}+\sum_{k=0} \frac{1}{100 (k^{2}+1)} (2^{-k}\cdot )2^k\epsilon^2
\end{align*}
Therefore, there exists some $k\geq 0$ such that
\begin{align*}
|\{i:2^{k-1}\epsilon^2\leq ||p_i-q_i||_2^2\}|>\frac{1}{2^k\cdot 100(k^{2}+1)}m.
\end{align*}
Actually, there is always such a $k\leq  k_0=\log_2 m$: If we find some $k>k_0$ with the above property, then the above property is also true for $k_0$,
\begin{align*}
&|\{i:2^{k-1}\epsilon^2\leq ||p_i-q_i||_2^2\}|>\frac{1}{2^{k}\cdot 100(k^{2}+1)}m.\\
\Rightarrow &|\{i:2^{k-1}\epsilon^2\leq ||p_i-q_i||_2^2\}|\geq 1\\
\Rightarrow &|\{i:2^{k_0-1}\epsilon^2\leq ||p_i-q_i||_2^2\}|\geq |\{i:2^{k-1}\epsilon^2\leq ||p_i-q_i||_2^2\}| \geq 1 
\geq\frac{1}{2^{k_0}\cdot 100(k_0^{2}+1)}m.
\end{align*}
Here, the probability of selecting some $i$ with this property is at least
\begin{align*}
1-(1-\frac{1}{100\cdot 2^k\times (k^{2}+1)})^{2^k\times (k^{2}+1) \times  L}\geq 1-\mathcal{O}(e^{-L/100}).
\end{align*}
After this, the corresponding tester will return "No" with high probability.

The sample complexity of this algorithm is
$$
\sum_{k=0}^{k_0} 2^k \cdot (k^{2}+1) \times \mathcal{O}(\frac{1}{2^{k-1}\epsilon^2} \times \log (L^2\cdot 6^k))=\mathcal{O}(\frac{\log^4 m}{\epsilon^2}). \footnote{Here, we have not attempted to optimize the exponent of $\log m$. It is direct to obtain $\mathcal{O}(\frac{\log^{3+\mu} m}{\epsilon^2})$ for any $\mu>0$ by selecting $2^{k}\cdot k^{1+\mu}\cdot C$ 
uniformly random elements $1\leq i\leq m$.}
$$
Although it looks that in the left handside  $2^{k_0} \cdot {k_0}^2>>m$, but this term  $2^{k_0} \cdot {k_0}^2$ appears only if $2^{k-1}\epsilon^2<1$.

This bound is tight upto the $\mathrm{poly}(\log m)$ factor: Even for the case $m=1$, $\frac{\log(\frac{1}{\delta})}{\epsilon^2}$ samples are needed to achieve successful probability $1-\delta$.

Back to the \textit{quantum identity testing} problem: Suppose we are given two $n$-qubit quantum states $\rho$ and $\sigma$, and we guarantee that either $\rho=\sigma$ or $||\rho-\sigma||_1>\epsilon$, and the goal is to distinguish them.

Without loss of generality, we can let
\begin{align*}
\rho=\frac{\sum_{P}\alpha_P P}{2^n}\\
\sigma=\frac{\sum_P \beta_P P}{2^n},
\end{align*}
where in the summation $P$ ranges over $\{\sigma_I,\sigma_X,\sigma_Y,\sigma_Z\}^{\otimes n}$. 
Then,
\begin{align*}
||\rho-\sigma||_2^2=\frac{\sum_P (\alpha_P-\beta_P)^2}{2^n}\geq \frac{||\rho-\sigma||_1^2}{2^n}.
\end{align*}
Therefore, if $\rho=\sigma$, $\alpha_i=\beta_i$ for all $i$.

If $||\rho-\sigma||_1>\epsilon$, we have
\begin{align*}
\sum_i (\alpha_P-\beta_P)^2\geq \epsilon^2.
\end{align*}
If we measure $\rho$ and $\sigma$ for Pauli $P_i$, the output of $\rho$ would be a sample of probability distribution $p_P=(\frac{1+\alpha_P}{2},\frac{1-\alpha_P}{2})$, and the output of $\sigma$ would be a sample of probability distribution $q_P=(\frac{1+\beta_P}{2},\frac{1-\beta_P}{2})$. That is
\begin{align*}
||p_P-q_P||_2^2=\frac{ (\alpha_P-\beta_P)^2}{2}.
\end{align*}
Therefore, to solve the \textit{quantum identity testing} problem, we only need to solve the ``Identity testing for distribution collections'' in a query model with $m=4^n$ and $\frac{\epsilon^2}{2\cdot 4^n}$. It can be solved using 
\begin{align*}
\mathcal{O}(\frac{n^4\cdot 4^n}{\epsilon^2})
\end{align*}
number of samples.

\subsection{Lower bound}
We use the quantum mixedness problem testing problem, a special case of Problem \ref{identity} by letting $\rho=\frac{I_{2^n}}{2^n}$, to provide lower bound.

If we regard Pauli measurement corresponding to $Q=Q_{+}-Q_{-} \in\{\sigma_I,\sigma_X,\sigma_Y,\sigma_Z\}^{\otimes n}$ as a two-outcome measurement $\{Q_{+},Q_{-}\}$, we can observe that $\Omega(\frac{4^n}{\epsilon^2})$ copies of $\sigma$ are necessary to solve the quantum mixedness testing. To see this, for any $P\in \{\sigma_I,\sigma_X,\sigma_Y,\sigma_Z\}^{\otimes n}$, we let
\begin{align*}
\sigma_P=\frac{I_{2^n}+\epsilon P}{2^n}
\end{align*}
Then,
\begin{align*}
||\sigma_P-\frac{I_{2^n}}{2^n}||_1=\epsilon
\end{align*}
Of course, to distinguish $\sigma_P$ from $\frac{I_{2^n}}{2^n}$, at least $\frac{1}{\epsilon^2}$ copies are needed to be measured in Pauli measurement corresponding to $P$.

Therefore, to distinguish $\frac{I_{2^n}}{2^n}$ from the uniform distribution over $\sigma_P$s, we need at least
\begin{align*}
(4^m-1)\frac{1}{\epsilon^2}={\Omega}(\frac{4^n}{\epsilon^2})
\end{align*}
copies.

Adaptively chosen Pauli measurement would not provide any advantage here.

As recently noticed in \cite{yu2020sample}, Pauli measurements is usually a $2^n$ outcome measurements. One can obtain a lower bound $\Omega(\frac{3^n}{\epsilon^2})$ by using 
\begin{align*}
\sigma_P=\frac{I_{2^n}+\epsilon P}{2^n}
\end{align*}
for any $P\in \{\sigma_X,\sigma_Y,\sigma_Z\}^{\otimes n}$.
\section{Acknowledgments}
This work was supported by DE180100156.

\bibliography{opt-tomo}

\end{document}